\documentclass[preprint,3p,times, nopreprintline]{elsarticle}
\usepackage{epstopdf}
\usepackage{graphicx}
\usepackage{amssymb}
\usepackage{tabularx}
\usepackage{subcaption} 
\usepackage{amsmath}
\usepackage{pifont}
\usepackage{supertabular}
\usepackage{eurosym}
\usepackage{xcolor} 
\usepackage{mathtools}
\usepackage{multirow}

\usepackage{lipsum}
\usepackage{multicol}

\usepackage{appendix}
\usepackage{booktabs}
\usepackage{array}

\usepackage{amsthm}

\newtheorem{definition}{Definition}

\usepackage[ruled]{algorithm2e}
\usepackage{algpseudocode}
\usepackage{float}
\usepackage{centernot}

\usepackage{natbib}
\usepackage[draft=False]{hyperref} 
\hypersetup{
     colorlinks=true,
     linkcolor=blue,
     filecolor=blue,
     citecolor=black,      
     urlcolor=cyan,
     }

\usepackage[acronym,toc]{glossaries}

\makeglossaries

\begin{document}

\newcommand{\rreh}{\gls{RREH}\xspace}
\newcommand{\rrehs}{\gls{RREH}s\xspace}
\newcommand{\pccc}{\gls{PCCC}\xspace}
\newcommand{\dac}{\gls{DAC}\xspace}
\newcommand{\cdo}{\gls{cdo}\xspace}
\newcommand{\capex}{\gls{CAPEX}\xspace}
\newcommand{\opex}{\gls{OPEX}\xspace}
\newcommand{\wacc}{\gls{WACC}\xspace}
\newcommand{\gwp}{\gls{gwp}\xspace}
\newcommand{\ptg}{\gls{ptg}\xspace}
\newcommand{\esc}{\gls{esc}\xspace}
\newcommand{\ch}{CH$_4$\xspace}
\newcommand{\hvdc}{\gls{hvdc}\xspace}

\begin{frontmatter}
    \author[ULIEGE]{Victor Dachet\corref{DACHET}}\ead{victor.dachet@uliege.be}
    \author[ULIEGE]{Antoine Dubois}
    \author[ULIEGE]{Bardhyl Miftari}
    \author[ULIEGE]{Raphaël Fonteneau}
    \author[ULIEGE]{Damien Ernst}
    \address[ULIEGE]{Department of Computer Science and Electrical Engineering, ULiège, Liège, Belgium}
    \cortext[DACHET]{Corresponding author}
    
    \title{
    Remote Renewable Energy Hubs: a Taxonomy
    }
    
    \begin{abstract}
    %
    %
    %
    %
    %
    Serving the energy demand with renewable energy is hindered by its limited availability near load centres (i.e. places where the energy demand is high). To address this challenge, the concept of Remote Renewable Energy Hubs (RREH) emerges as a promising solution. RREHs are energy hubs located in areas with abundant renewable energy sources, such as sun in the Sahara Desert or wind in Greenland. In these hubs, renewable energy sources are used to synthetise energy molecules. To produce specific energy molecules, a tailored hub configuration must be designed, which means choosing a set of technologies that are interacting with each other as well as defining how they are integrated in their local environment. The plurality of technologies that may be employed in RREHs results in a large diversity of hubs. In order to characterize this diversity, we propose in this paper a taxonomy for accurately defining these hubs. This taxonomy allows to better describe and compare designs of hubs as well as to identify new ones. Thus, it may guide policymakers and engineers in hub design, contributing to cost efficiency and/or improving local integration.
    
    
    
    \end{abstract}
\end{frontmatter}

\section{Introduction} \label{sec:introduction}

To decarbonize their energy supply, load centers, i.e., geographical zones with high energy demand, will have to rely on large amounts of renewable energy.
Nevertheless, renewable energy produced locally in those load centers, such as Belgium, may be insufficient to cover all energy needs for various reasons \cite{berger2020role}, such as:
(i) limited space to install renewable energy assets notably due to factors such as strong urbanization or geographical constraints and
(ii) low-quality renewable energy sources.
%
To address these limitations, Remote Renewable Energy Hubs (RREHs), i.e., energy hubs situated away from those load centers where renewable energy is abundant, offer a solution to the lack of local renewable resources.
RREHs have spurred significant research on possible hub development around the globe \cite{berger2020role, dachet2023co2, hashimoto1999global}.
They can rely on power-to-X technologies that present a dual advantage \cite{gotz2016renewable, o2016energy, munster2020sector}.
They offer a CO$_2$-neutral solution to meet energy demand and a means of storing energy generated by renewable sources \cite{blanco2018review}.

Different models of RREHs have been proposed in the literature.
\citet{hashimoto1999global} proposed a RREH, where CH$_4$ was produced in Egypt to deliver methane to Japan. 
However, no techno-economic analysis of the supply chains was carried out.
Then, \citet{fasihi2015economics} and \citet{fasihi2017longterm} proposed a techno-economic analysis for the production of CH$_4$ in Northern Africa and delivery in Finland. 
In \citet{agora2018future}, the authors investigate a similar supply chain with delivery in Germany. 
\citet{berger2021remote} proposed a techno-economic analysis of a supply chain between the Sahara desert in Algeria and Belgium as a load centre. 
\citet{dachet2023co2} performed a techno-economic analysis of introducing a loop involving the export of CO$_2$ from Belgium to the RREH defined in \citet{berger2021remote} as proposed by \citet{hashimoto1999global}. 
They highlight the potential of post-combustion carbon capture (PCCC) in Belgium to valorise CO$_2$ in a RREH.
They also investigated another RREH located in Greenland, which has also been proposed as an energy hub for Europe with complementary wind regimes \cite{radu2019complementarity, radu2022assessing}. 
\citet{fonder2023synthetic} explore the same idea with PCCC technologies in complement of DAC devices to produce CO$_2$ in a RREH located in Morocco. 
\citet{verleysen2023where} have studied a case of RREH exporting NH$_3$ from Morocco to Belgium.
They assessed that the cost of a RREH would be lower than an ammonia hub in Belgium. 
Moreover, uncertainties on CAPEX and OPEX technology costs have also been taken into account in \cite{verleysen2023where}.
\citet{Larbanois2023MultiCarriers} compared four energy vectors, namely hydrogen (H$_2$), ammoniac (NH$_3$), methanol (CH$_3$OH), and methane (CH$_4$) synthesized in Algeria to meet an energy demand in Belgium. 
%


On top of the interest from the scientific community, several industrial RREH projects are being developped. For example, BP aims to establish an RREH in northern Australia to export e-ammonia, primarily to Japan and South Korea \cite{BP_AustralianRenewableEnergyHub}. Another example is CMB.Tech, a Belgian maritime group that plans to develop an RREH in Namibia in partnership with the Namibian government to export e-ammonia as fuel for their vessels \cite{CleanergySolutionsNamibia}. One reason industrial players are interested in RREHs is the policy targets set to drive e-fuel demand. For instance, the European Union aims to decarbonize maritime and aviation transport by replacing part of their fossil fuels with e-fuels \cite{FuelEUMaritime, ReFuelEUAviation}. On a larger scale, according to the International Renewable Energy Agency (IRENA), in their scenario for limiting global warming to 1.5°C, e-fuels could contribute up to 12\% of the worldwide CO$_2$ emissions reduction \cite{IRENA2023}. RREHs producing e-fuels can help to achieve these worldwide CO$_2$ emmission reduction targets.

It is worth noticing that the global grid approach \cite{chatzivasileiadis2013global, liu2015global, yu2019global}, is closely related to the concept of RREH.
Indeed, a global grid would allow to collect renewable energy where it is the most abundant and to meet an electricity demand far away from the harvested zone.
%
%
The DESERTEC project \cite{samus2013assessing} that repatriates renewable energy through electric cables to Europe is an example of a project to develop a global grid. 
The global grid approach can be seen as the interconnection of load centres and RREHs around the world with electricity as an export commodity.


There exists a combinatorial number of possibilities to design a hub due to all the technologies that can be considered. This implies a large diversity of hubs, making it difficult to compare different hubs, even if they are located in the same place or aimed at producing the same energy vectors. Indeed, RREHs producing the same energy vectors may also differ in their import commodities which complexifies again their comparison and their description.
Moreover, in the literature, there is often a lack of clarity regarding whether the valorization of local opportunities and by-products inherent to this type of infrastructure is taken into account. To address these problems, we propose a taxonomy relying on a  mathematical sets formalization to better characterize RREHs. This taxonomy aims to provide a tool for describing and comparing RREH models. Moreover, it may be useful for those seeking to improve the design of new hubs. This taxonomy could also offer new prospects for enhancing the local integration of these hubs. To the best of the authors' knowledge, this taxonomy is the first to formally address the design phase of RREHs, which has not been previously discussed or structured within a given framework.


We define the concept of RREH in the next section. In \autoref{sec:taxonomy}, we present our taxonomy and illustrate it with an academic example. In \autoref{sec:discussion}, we apply the taxonomy to two hubs studied in the literature, demonstrating its effectiveness in describing and comparing hubs. In \autoref{sec:design_and_integration}, we propose a practical procedure for using the taxonomy to identify new hubs. Finally, \autoref{sec:conclusion} concludes the paper.

\section{Definition and Concept}\label{sec:definition}

\begin{definition}
    An \textbf{energy hub} integrates input and output of commodities, conversion, and storage functionalities, enabling coupling between different energy systems.
    
    An energy hub can also  encompass production/consumption units, and transportation infrastructure, allowing the exchange of multiple energy carriers.
\end{definition}

This definition has been adapted from \cite{geidl2007integrated}, omitting references to sustainable energy.
In both definitions, an offshore wind farm or an onshore solar farm are considered as energy hubs whereas this new definition also encompasses an offshore petroleum platform.
%
In this regard, we distinguish the concept of energy hub from the concept of a \textit{ renewable energy hub} which is defined in the following way:

\begin{definition}\label{def:REH}
    A \textbf{renewable energy hub} is an energy hub that relies on renewable energy sources for energy production.
\end{definition}

Energy hubs can be located anywhere on Earth.
However, it makes sense to distinguish energy hubs that are located in remote areas, far from the main centres of population.
Notably due to their remote nature, numerous geographical zones with high renewable energy density are yet to be exploited.
Additionally, setting energy hubs away from population centres allows the deployment of large infrastructures without impacting the lives of citizens. 
This leads us to define and focus on what we denote as Remote Renewable Energy Hubs.

\begin{definition}\label{def:RREH}
    A \textbf{Remote Renewable Energy Hub} (RREH) is a renewable energy hub located in a remote area.
\end{definition}

In this definition, we acknowledge that remoteness is a subjective notion, but this is an intentional choice: it allows for the inclusion under the taxonomy described in the following sections of a broad range of energy hubs. Moreover, it underscores the necessity of transportation for the energy produced.

\section{Taxonomy}\label{sec:taxonomy}

\begin{figure*}[h]
    \centering
    \includegraphics[width=0.5\textwidth]{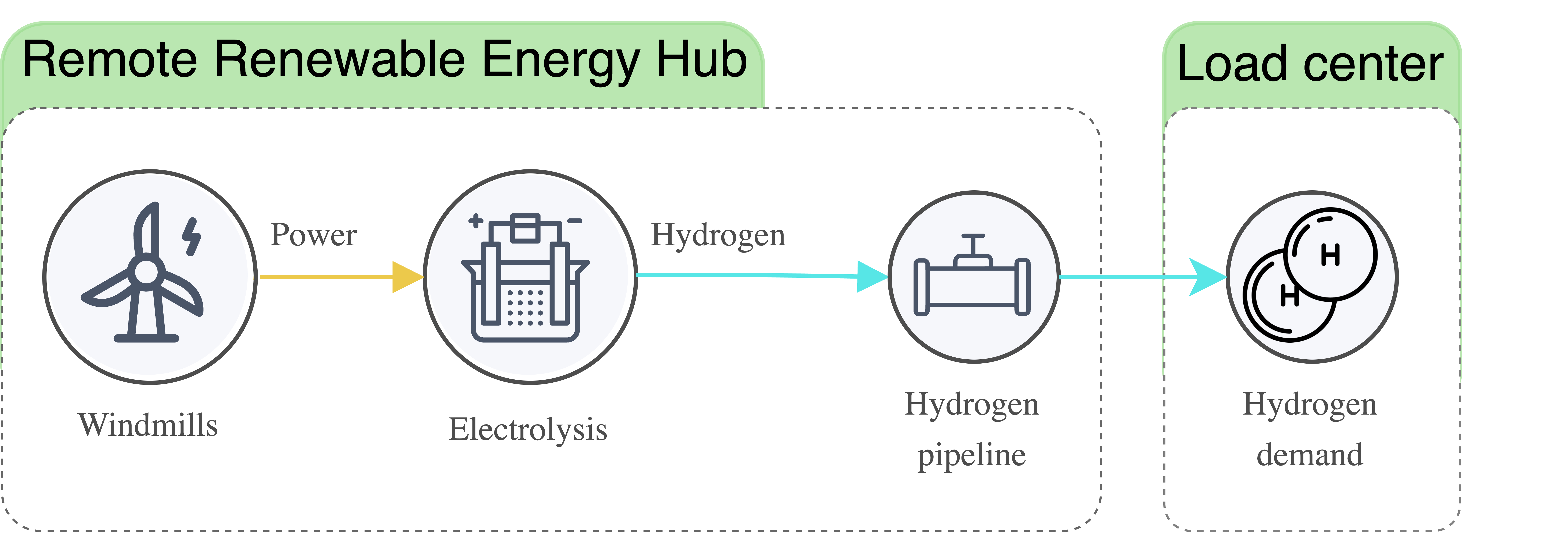}
    \caption{RREH located in Greenland exporting H$_2$ towards North America.}
    \label{fig:GL_example}
\end{figure*}

This section formally defines the taxonomy, and, in particular, explains the purpose of each of its components in the context of a RREH. The first part introduces conceptual mathematical elements that are not related to any particular hub in \autoref{subsec:gen_math_ele}. The second part, \autoref{subsec:ele_taxo_tailored}, uses these conceptual mathematical elements to derive elements of the taxonomy tailored to any specific hub. To illustrate this, a hypothetical RREH in Greenland is considered. This RREH generates renewable electricty from wind turbines that is used to power an electrolyzer. The H$_2$ produced by the electrolyzer is subsequently transported via a pipeline to Iceland. A visual representation of this hypothetical hub is provided in \autoref{fig:GL_example}. Finally, \autoref{fig:schematic} summarizes the components relationships.

\subsection{Conceptual mathematical elements}\label{subsec:gen_math_ele}

Before instantiating the mathematical elements that constitute an RREH, four conceptual mathematical elements are formally defined, namely $\mathcal{C}, \mathcal{L}, \mathcal{T}$ and $\mathcal{H}$. These are defined as follows:

\begin{itemize}
    \item $\mathcal{C}$: the set of all commodities that can be exchanged between technologies typically composed of chemical components, electricity and heat. 

    \item $\mathcal{L}$: the set of locations and their specificities in terms of renewable energy potential and energy demand. An element $l \in {\mathcal L}$ is characterised by the triplet  $l = (L, P_l, D_l)$ where $L$ is a location and $P_l$ and $D_l$ are respectively the renewable potential and the demand associated with the location $L$. A location characterised by a high demand and low renewable potential is referred to as a load centre. 
    
    \item $\mathcal{T}$: the set of all possible technologies. A technology $t \in \mathcal{T} $, associated with a name $n$, can be seen as a function processing input commodities $\mathcal{C}_{t}^{in} \subseteq \mathcal{C}$ and converting them into output commodities $\mathcal{C}_{t}^{out} \subseteq \mathcal{C}$, represented as a 3-tuple $t = \left( n, \mathcal{C}_{t}^{in}, \mathcal{C}_{t}^{out} \right)$. Included in the set of technologies $\mathcal{T}$, there are three specific technologies: 
    \begin{enumerate}[(i)]
        \item $t^{im} = \left( \text{import}, \emptyset, \mathcal{C}_{t^{im}}^{out} \right)$ models the imported commodities;
        \item $t^{ex} = \left( \text{export}, \mathcal{C}_{t^{ex}}^{in}, \emptyset \right)$ models the exported commodities;
        \item $t^{op} = \left( \text{opportunity}, \mathcal{C}_{t^{op}}^{in}, \emptyset \right)$ models the locally exploited commodities;
    
    \end{enumerate}
        
    
    \item $\mathcal{H}$: the set of possible flows of commodities. In this set each element $h = \left(c, \mathcal{T}^{out}, \mathcal{T}^{in} \right) \in \mathcal{H}$ is made of three components: a commodity $c \in \mathcal{C}$, a set of technologies $\mathcal{T}^{out} \subseteq \mathcal{T} $ that output the commodity $c$ and a set of technologies $\mathcal{T}^{in} \subseteq \mathcal{T}$ that have as input the commodity $c$ coming from the technologies in    $\mathcal{T}^{out}$. Therefore, $h$ denotes a flow of commodity $c$ from each technology contained in $\mathcal{T}^{out} $ to each technology contained in $\mathcal{T}^{in}$.
    One can observe that the edges follow the convention of hyperedges in this work. This simplifies the taxonomy's readability by reducing the number of edges that need to be defined. Specifically, a single hyperedge represents $| \mathcal{T}^{out} | \times | \mathcal{T}^{in} |$ edges where $|\mathcal{X}|$ corresponds to the cardinality of the set $\mathcal{X}$.
\end{itemize}

\subsection{Elements of the taxonomy tailored to a specific hub}\label{subsec:ele_taxo_tailored}

Now that the conceptual mathematical elements have been introduced, any RREH $r$ can be formalised as a 7-tuple $r= \left(\mathcal{L}_r, \mathcal{G}_{r}, \mathcal{C}_{r}, \mathcal{E}_r, \mathcal{I}_r, \mathcal{B}_r, \mathcal{O}_r \right)$ can be characterised by its components: 

\begin{itemize}
    \item $\mathcal{L}_r \subseteq \mathcal{L}$: a set of locations associated  with the technologies in the hub. Indeed, to model a RREH, a location or several must be identified. The location determines the RREH's potential in terms of resource availability such as solar, wind, hydropower or geothermia. Moreover, the geographical location of the RREH will influence its competitiveness in the closest load centres. 
    
    In the example provided in \autoref{fig:GL_example}, the set of locations includes only one location in Greenland that has a high renewable energy potential in wind and a low energy demand due to the low density popultation. Therefore, $$\mathcal{L}_r = \{l_1\},$$ 
    where 
    
    $$l_1 = (\text{Greenland}, \text{high wind potential}, \text{low energy demand}).$$ 

    \item $\mathcal{G}_{r} = (\mathcal{T}_{r}, \mathcal{H}_{r})$: a graph that mathematically formalises the technologies and commodity flows, represented as $\mathcal{T}_{r} \subseteq \mathcal{T}$ and $\mathcal{H}_r \subseteq \mathcal{H}$. These technologies $\mathcal{T}_{r}$ are situated in locations depicted in $\mathcal{L}_r$. The relationships between these components form a graph structure, with $\mathcal{T}_{r}$ denoting the nodes and $\mathcal{H}_{r}$ representing the edges. Defining $\mathcal{G}_{r}$ is called designing a RREH. Therefore, designing a RREH consists of finding the technologies (\textit{i.e.} the nodes in $\mathcal{T}$) that compose the RREH and the flows of commodities between these technologies (\textit{i.e.} the edges in $\mathcal{H}$). 

    In our Greenland example (cfr. \autoref{fig:GL_example}), the graph of technologies is defined as: 

    \begin{tabularx}{0.5\textwidth}{p{0.07\textwidth}X}
        $\mathcal{T}_{r} = \{$ &

        \begin{description}
            \item[] $(\text{Wind}_{l_1}, \{ \emptyset \}, \{ \text{electricity} \}),$
            \item[] $(\text{import}_{l_1}, \{ \emptyset \}, \{ \text{H}_2\text{O} \}),$
            \item[] $(\text{electrolyzer}_{l_1}, \{ \text{electricity}, \text{H}_2\text{O} \}, $
            \item[] $\{ \text{H}_2, \text{O}_2 \}),$
            \item[] $(\text{export}_{l_1}, \{ \text{H}_2 \}, \{ \emptyset \})$
        \end{description}
        
        $\}$\\
        $\mathcal{H}_{r} = \{$ & 
        
        \begin{description}
            \item[] $ (\text{Electricity}, \{\text{Wind}_{l_1}\}, \{\text{electrolyzer}_{l_1} \})$,
            \item[] $(\text{H}_2\textbf{O}, \{ t^{im}_{l1} \}, \{\text{electrolyzer}_{l_1} \}),$ 
            \item[] $(\text{O$_2$}, \{ \text{electrolyzer}_{l_1} \}, \emptyset),$ 
            \item[] $(\text{H}_2, \{\text{electrolyzer}_{l_1} \}, \{t^{ex}_{l1} \}) $
        \end{description}
        $\}.$
         \\
    \end{tabularx}

    \item $\mathcal{C}_{r} \subseteq \mathcal{C}$: the set of commodities within the RREH. $\mathcal{C}_{r}$ can be derived from $\mathcal{G}_{r}$. 

    In our Greenland example (cfr. \autoref{fig:GL_example}), 
    
    $\mathcal{C}_{r} = \{ \text{electricity}, \text{H}_2\text{O}, \text{H}_2, \text{O}_2 \}$.

    \item $\mathcal{E}_r \subseteq \mathcal{C}_{r}$: the set of exported commodities. These commodities are input in $t^{ex} \in \mathcal{T}_{r}$ that are exported to a load centre (\textit{i.e.} to a location $L^{'} \notin \mathcal{L}_r$) in a selected carrier, such as electricity, liquid or gaseous hydrogen. The so-called design process entails determining the necessary processes for producing the exported commodities in $\mathcal{E}_r$.

    In our Greenland example (cfr. \autoref{fig:GL_example}), 
    
    $\mathcal{E}_r = \{ \text{H}_2 \}$.

    \item $\mathcal{I}_r \subseteq \mathcal{C}_{r}$: the set of imported commodities. These commodities are outputted by $t^{im} \in \mathcal{T}_{r}$.
    The design of the RREH may also be influenced by the availability of imported commodities in the hub (\textit{i.e.} to a location in $\mathcal{L}_r$). 
    Indeed, it is not necessary to produce all required commodities in close proximity to the RREH. For instance, as demonstrated by \citet{dachet2023co2}, importing CO$_2$ from the load centre rather than relying on DAC technologies can lead to a reduction in the total system cost.
    In our Greenland example (cfr. \autoref{fig:GL_example}),
    $\mathcal{I}_r = \{ \text{H}_2\text{O} \}$. \\

    \item $\mathcal{B}_r \subseteq \mathcal{C}_{r}$: the set of byproduct commodities. Among all the commodities within the RREH (\textit{i.e.} in $\mathcal{C}_{r}$), some are byproducts that are never used in any process. More specifically, these byproducts are commodities output by a technology $t \in \mathcal{T}_{r}$ but not involved into any edge composed of this commodity output by $t$ and included in the input set of any technology $t' \in \mathcal{T}_{r}$. These byproduct commodities are never used within the RREH however they could be valorised by re-designing the RREH to input them into an existing technology of the hub or a new technology not already considered. Another possibility is to valorize those commodities in a market. 

    In our Greenland example (cfr. \autoref{fig:GL_example}),
    
    $\mathcal{B}_r = \{ \text{O}_2 \} $. 

    \item $\mathcal{O}_r \subseteq \mathcal{C}_r$: the set of locally exploited opportunities. This set entails the commodities that are input in $t^{opportunity} \in \mathcal{T}_{L^{'}}$ that are exported to meet a local demand (\textit{i.e.} to a location $L^{'} \in \mathcal{L}_r$). In fact, all the commodities within the RREH can represent opportunities for local development, thereby facilitating local integration of the RREH. 

    In our Greenland example (cfr. \autoref{fig:GL_example}), no produced commodity has been used to meet a local demand, hence
    $\mathcal{O}_r = \emptyset  $. 
    
\end{itemize}

\begin{figure}
    \centering
    \includegraphics[width=0.5\textwidth]{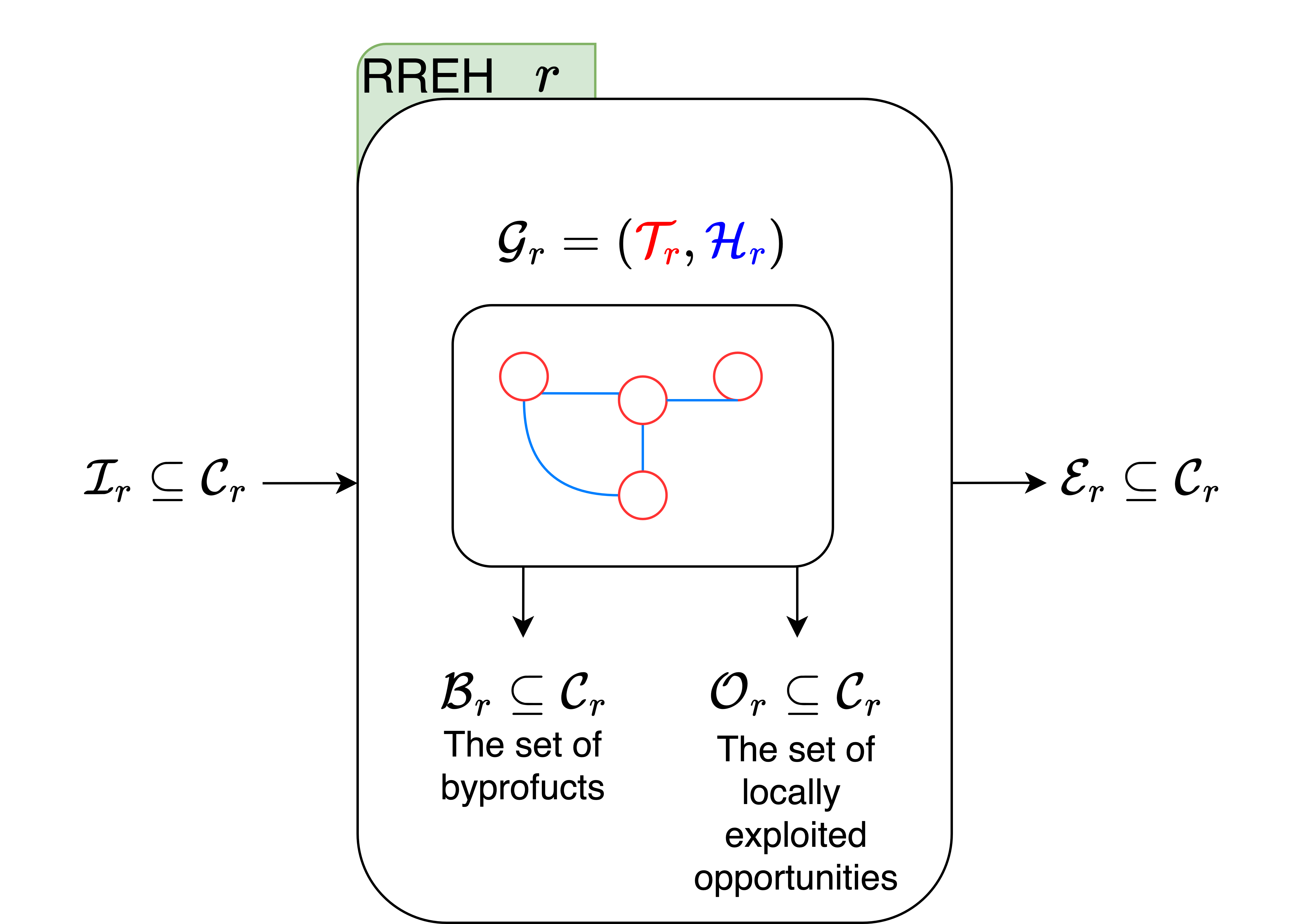}
    \caption{Schematic view of a RREH with the different sets which define it.}
    \label{fig:schematic}
\end{figure}

A schematic view of these sets and graph is given in \autoref{fig:schematic} where the set of imported commodities $\mathcal{I}_r$ is processed by the graph of technologies $\mathcal{T}_{r}$ and the commodities flows  $\mathcal{H}_{r}$ in the RREH producing local opportunities $\mathcal{O}_r$ and by-products $\mathcal{B}_r$ and exporting commodities included in the set of exports $\mathcal{E}_r$.

\section{Instantiation}\label{sec:discussion}

In this section, the taxonomy will be exemplified using a hub located in Algeria, as studied in the literature \cite{berger2020role}. This will highlight how the taxonomy can be used to easily describe hubs. Then, a second example from the literature \cite{Larbanois2023MultiCarriers} will be presented using the taxonomy to demonstrate how the taxonomy can ease the comparison between them.

\begin{figure*}[h]
    \centering
    \includegraphics[width=0.9\textwidth]{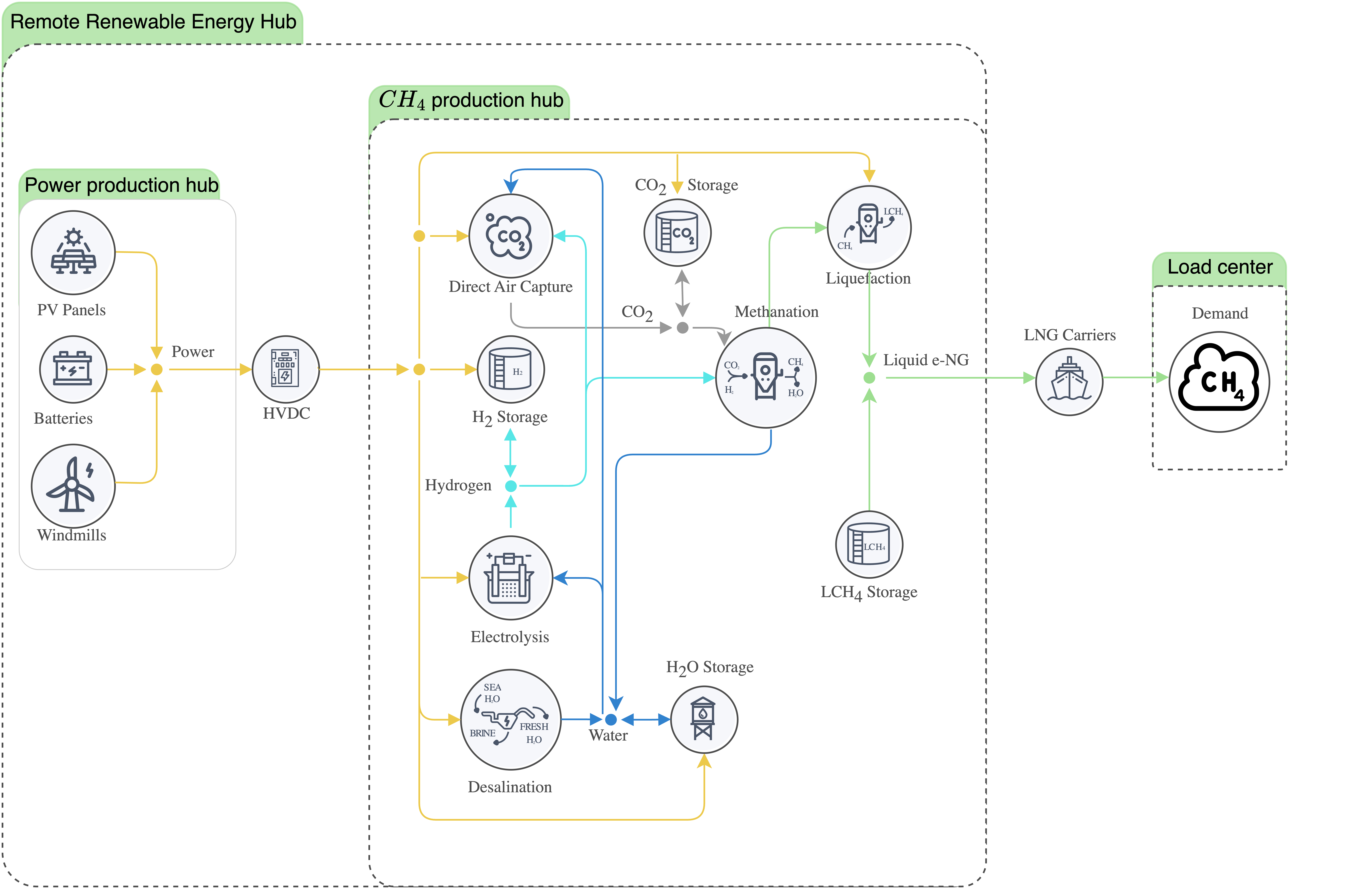}
    \caption{RREH located in Algeria exporting CH$_4$ to a load center situated in Belgium, adapted from \cite{berger2021remote}. }
    \label{fig:RREH_CH4}
\end{figure*}

The first RREH $r_1$, coming from \cite{berger2021remote}, is composed of renewable energy production (solar and wind) that powers an electrolyzer to produce hydrogen. From this hydrogen and CO$_2$ captured via DAC, a methanation process produces methane (CH$_4$) that is liquefied and exported by boats to Belgium. This RREH is divided into two connected parts: one located in the Sahara desert for harnessing renewable energy and a second one located on the Algerian coast responsible for the synthesis and export of methane to Belgium. These two parts are connected via a High Voltage Direct Current (HVDC) link. In the taxonomy, this RREH would be expressed as:

\begin{itemize}
    \item $\mathcal{L}_{r_1} = \{ l_1, l_2, l_3\}\\
            \text{ where} \\
            l_1 = (\text{Sahara desert}, \\ \text{high renewable potential}, \text{low demand}), \\
            l_2 = (\text{Algerian coast}, \text{high renewable potential}, \\ \text{medium demand}) \}, \\
            l_3 = (\text{From Sahara desert to the coast}, \\ \text{high renewable potential}, \text{low demand})$ 
    \item $\mathcal{G}_{r_1} = (\mathcal{T}_{r_1}, \mathcal{H}_{r_1})$ is represented in \autoref{fig:RREH_CH4} and comprehensively detailed in \autoref{tab:taxonomy}.  
    \item $\mathcal{C}_{r_1} = \{ \text{electricity}, CH_4(g), CH_4(l), H_2, H_{2}O, CO_2, O_2, \text{heat} \}$
    \item $\mathcal{E}_{r_1} = \{ CH_4 \}$ 
    \item $\mathcal{I}_{r_1} = \{ \text{sea water} \}$ 
    \item $\mathcal{B}_{r_1} = \{ O_2, \text{heat} \} $
    \item $\mathcal{O}_{r_1} = \{  \} $.
\end{itemize}

The comprehensive description is available in \autoref{tab:taxonomy}. 

The second RREH $r_2$, coming from \cite{Larbanois2023MultiCarriers}, is composed of renewable energy production (solar and wind) that powers an electrolyzer to produce hydrogen. From this hydrogen and nitrogen (N2) produced via an Air Separation Unit (ASU), an Haber-Bosch process synthesizes ammoniac (NH$_3$) in liquid form that is exported by boats to Belgium. This RREH is located and divided at the same locations than the other ones. 

The expression of this second hub is avalaible in \autoref{tab:RREH_comparison} and the comprehensive description is available in \autoref{tab:taxonomy-example2}. The \autoref{tab:RREH_comparison} also highlights the key differences between these two hubs. As an example, the main differences between these hubs, besides the exported commodity, are the byproducts. The hub exporting CH$_4$ does not produce Argon whereas the hub exporting NH$_3$ does. 



\begin{table*}[h!]
    \centering
    \caption{Characteristics and differences of Remote Renewable Energy Hubs \( r_1 \) and \( r_2 \).}
    \label{tab:RREH_comparison}
    \begin{tabular}{|c|p{6.5cm}|p{4cm}|p{4cm}|}
        \hline
         & $r_1$ & $r_2$ & Differences \\
        \hline
        \(\mathcal{L}_r\) & 
        \( \begin{aligned}
            &\{ l_1, l_2, l_3 \} \\
            & l_1 = (\text{Sahara desert}, \text{high renewable potential},  \\
            & \text{low demand}), \\
            & l_2 = (\text{Algerian coast}, \text{high renewable potential}, \\
            & \text{medium demand}), \\
            & l_3 = (\text{From Sahara desert to the coast},  \\
            & \text{high renewable potential}, \text{low demand}) \\
          \end{aligned} \) & 
        Same as \( \mathcal{L}_{r_1} \) & - \\
        \hline
        \(\mathcal{G}_r\) & 
        \((\mathcal{T}_{r_1}, \mathcal{H}_{r_1})\) (See Figure~\ref{fig:RREH_CH4} and Table~\ref{tab:taxonomy}) & 
        \((\mathcal{T}_{r_2}, \mathcal{H}_{r_2})\)
        (See Figure~\ref{fig:RREH_NH3} and Table~\ref{tab:taxonomy-example2}) & Different technological and hub structures \\
        \hline
        \(\mathcal{C}_r\) & 
        $\begin{aligned}
        & (\{\text{electricity}, CH_4(g), CH_4(l), H_2, H_2O, 
        CO_2, \\
        & O_2, \text{heat}\}) 
        \end{aligned}$ & 
        $\begin{aligned}
            &(\{\text{electricity}, NH_3, H_2, H_2O, \\
            &N_2, O_2, \text{heat}, Ar\}) \\
        \end{aligned}$ & Different energy molecule: CH\textsubscript{4} vs. NH\textsubscript{3}, plus presence of Ar in \( r_2 \) \\
        \hline
        \(\mathcal{E}_r\) & 
        \(\{ CH_4 \}\) & 
        \(\{ NH_3 \}\) & Energy export differs \\
        \hline
        \(\mathcal{I}_r\) & 
        \(\{\text{sea water}\}\) & 
        Same as \( \mathcal{I}_{r_1} \) & - \\
        \hline
        \(\mathcal{B}_r\) & 
        \(\{O_2, \text{heat}\}\) & 
        \(\{O_2, \text{heat}, Ar\}\) & Ar is present in \( r_2 \) but not in \( r_1 \) \\
        \hline
        \(\mathcal{O}_r\) & 
        \(\{\}\) & 
        Same as \( \mathcal{O}_{r_1} \) & - \\
        \hline
    \end{tabular}

\end{table*}

\begin{table*}[htbp]
    \small
    \centering
    \caption{Expression of the full taxonomy where the graph of technologies and flows of commodities $\mathcal{G}_{r}$ is described by its set of nodes and edges $(\mathcal{T}_{r}, \mathcal{H}_{r})$. }
    \begin{tabular}{|p{2cm}|p{12cm}|}
        \hline
        \textbf{Set} & \textbf{Description} \\
        \hline
        $\mathcal{L}_r$ & \begin{description} 
            \item $\{ l_1, l_2, l_3\}$
            \item $\text{ where} $
            \item $l_1 = (\text{Sahara desert},  \text{high renewable potential}, \text{low demand}),$ 
            \item $l_2 = (\text{Algerian coast}, \text{high renewable potential}, \text{medium demand}) \},$ 
            \item $l_3 = (\text{From Sahara desert to the coast}, \text{high renewable potential}, \text{low demand})$
        \end{description} \\
        \hline
        $\mathcal{T}_{r}$ & 
        
        \begin{description}
        \item $\{(PV_{l_1}, \{ \emptyset \}, \{\text{electricity} \}),$
        \item $(\text{Wind}_{l_1}, \{ \emptyset \}, \{\text{electricity} \}),$
        \item $(\text{Battery}_{l_1}, \{\text{electricity} \}), \{\text{electricity} \}),$
        \item $(\text{HVDC}_{l_3}, \{\text{electricity} \}, \{\text{electricity} \}),$
        \item $(\text{electrolyzer}_{l_2}, \{\text{electricity}, \text{H}_2\text{O} \}, \{\text{H}_2, \text{O}_2 \}),$
        \item $(\text{Desalination}_{l_2}, \{\text{electricity}, \text{sea water} \}, \{\text{H}_2\text{O} \}),$
        \item $(\text{H}_2\text{-Storage}_{l_2}, \{\text{electricity}, \text{H}_2 \}, \{\text{H}_2 \}),$
        \item $(\text{DAC}_{l_2}, \{\text{electricity}, \text{H}_2\text{O} \}, \{\text{CO}_2 \}),$
        \item $(\text{CO}_2\text{-Storage}_{l_2}, \{\text{electricity}, \text{CO}_2 \}, \{\text{CO}_2 \}),$
        \item $(\text{Methanation}_{l_2}, \{\text{CO}_2, \text{H}_2 \}, \{\text{CH}_4\text{(g)}, \text{H}_2\text{O} \}),$
        \item $(\text{CH}_4\text{-Liquefaction}_{l_2}, \{\text{electricity}, \text{CH}_4\text{(g)} \}, \{\text{CH}_4\text{(l)} \}),$
        \item $(\text{CH}_4\text{-Storage}_{l_2}, \{\text{CH}_4\text{(l)} \}, \{\text{CH}_4\text{(l)} \}),$
        \item $(\text{export}_{l_2}, \{\text{CH}_4\}, \{\emptyset \})\}$
    \end{description} \\
        \hline
        $\mathcal{H}_{r}$ & 
            \begin{description}
                \item $\{(\text{Electricity}, \{\text{Wind}_{l_1}, \text{Battery}_{l_1}, PV_{l_1}\}, \{\text{Battery}_{l_1},\text{HVDC}_{l_1} \}),$ 
                \item $(\text{Electricity}, \{\text{HVDC}_{l_1}\}, $  
                \item \hspace{0.8cm} $\{\text{Battery}_{l_2},\text{electrolyzer}_{l_2}, \text{Desalination}_{l_2}, \text{H$_2$-Storage}_{l_2},DAC_{l_2},\text{CO$_2$-Storage}_{l_2}\}),$
                
                \item $(\text{H$_2$O}, \{\text{H$_2$-Storage}_{l_2}, \text{Desalination}_{l_2}, \text{Methanation}_{l_2} \},$
                
                \item \hspace{0.8cm} $\{ \text{H$_2$-Storage}_{l_2}, \text{electrolyzer}_{l_2}, DAC_{l_2} \}  ),$
                
                \item $(\text{H$_2$}, \{\text{H$_2$-Storage}_{l_2}, \text{electrolyzer}_{l_2} \}, \{\text{H$_2$-Storage}_{l_2}, \text{Methanation}_{l_2}, DAC_{l_2} \}),$
                \item $(\text{CH$_4$(g)}, \{ \text{Methanation}_{l_2} \}, \{ \text{CH$_4$-Liquefaction}_{l_2}\}),$
                \item $(\text{CH$_4$(l)}, \{ \text{CH$_4$-Liquefaction}_{l_2}, \text{CH$_4$-Storage}_{l_2} \}, \{ \text{CH$_4$-Storage}_{l_2}, t^{ex} \}),$
                \item $(\text{O$_2$}, \{ \text{electrolyzer}_{l_2} \}, \emptyset),$
                \item $(\text{Heat}, \{ \text{electrolyzer}_{l_2} \}, \emptyset),$
                \item $(\text{Heat}, \{ \text{Methanation}_{l_2} \}, \emptyset),$
                \item $(\text{Heat}, \{ \text{CH$_4$-Liquefaction}_{l_2} \}, \emptyset)\}$
            \end{description} \\
        \hline
        $\mathcal{C}_{r}$ & $ \{ \text{Electricity}, CH_4(g), CH_4(l), H_2, H_2O, CO_2, O_2, \text{Heat} \}$ \\
        \hline $\mathcal{E}_r $ & $ \{ CH_4 \}$  \\
        \hline $\mathcal{I}_r $ & $ \{ \text{sea water} \}$  \\
        \hline $\mathcal{B}_r $ & $ \{ O_2, \text{Heat} \} $ \\
        \hline $\mathcal{O}_r $ & $ \{  \} $ \\
        \hline
    \end{tabular}
    
    \label{tab:taxonomy}
\end{table*}

\section{Design and Local Integration}\label{sec:design_and_integration}

In this section, we first propose, in \autoref{subsec:systematic_approach}, a systematic approach to guide the design process - specifically, identifying the technologies that constitute the RREH and the connections among them - as well as the local integration of a hub. Each step of this approach is then discussed with reference to existing literature on hubs. Then, in \autoref{subsec:example_prodcedure}, an example is given to illustrate how to use this approach in practice. 

\subsection{Systematic approach}\label{subsec:systematic_approach}

The approach is as follows: 
\begin{enumerate} 
    \item \textbf{Define Export Commodities}: identify the export set $\mathcal{E}_r$ to establish which commodities (e.g., methane, methanol, electricity) the RREH will produce for export.

    \item \textbf{Select locations}: from all the locations available worldwide, identify those suitable for energy harvesting and transport to define your set $\mathcal{L}_r$. 
    
    \item \textbf{Construct Technological Graph}: develop a potential technology graph $\mathcal{G}_{r}$ to describe the required technologies and commodities for producing the items in $\mathcal{E}_r$.
    
    \item \textbf{Consider Imports}: assess potential import commodities $\mathcal{I}_r$, especially those scarce locally but obtainable from elsewhere, to potentially reduce the number of technologies required to produce commodities within the RREH for its operation. 
    
    \item \textbf{Assess Byproducts}: evaluate byproducts $\mathcal{B}_r$ as potential resources that can be integrated into the RREH design to optimize commodity reuse and reduce operational costs.
    
    \item \textbf{Identify Local Opportunities}: use $\mathcal{C}_r$ that list available commodities and determine locally valuable ones. One may valorize these and add these to the set $\mathcal{O}_r$ to strengthen local integration.
    
    \item \textbf{Optimize}: based on the results of step 1-6, optimize the hubs based on your objective criteria and assess the results. If the results are not deemed to be satisfactory, repeat the different steps to identify another hub design.
\end{enumerate}

Step 1 may involve several considerations to determine which molecule to export, including which commodities are easiest to transport over a given distance and the infrastructure available in the load center to meet demand (e.g., CH$_4$ or H$_2$ networks).

Step 2, selecting locations, can be approached qualitatively by examining renewable energy potential maps or by identifying existing infrastructure, such as gas terminals, dedicated to the selected export molecule. Alternatively, it can be conducted more quantitatively using combinatorial optimization methods, such as those proposed by \citet{radu2022assessing}, to determine optimal locations for energy production.

Step 3 requires consulting the scientific literature to identify the chemical and physical processes necessary for producing a given export molecule. For instance, \cite{Larbanois2023MultiCarriers} proposes hubs exporting ammonia, hydrogen, or methanol, each requiring a unique design.

Step 4 may help reduce production costs. For example, \cite{dachet2023co2} demonstrates that importing CO$_2$ captured at the load center lowers the cost of methane synthesis compared to solely local production via DAC facilities.

Step 5 can also reduce operational costs, as shown by \citet{Dauchat2023HeatValoHubs}, who reuse process-generated heat and provide a detailed analysis of its valorization within an RREH, resulting in cost reductions. Additionally, this step may highlight the need for new components, such as Heat Recovery Steam Generators to reduce overall costs \cite{Dauchat2023HeatValoHubs}.


Step 6 helps to prevent project failure, as noted in \cite{courrier2024tunisie}. Additionally, it can decrease the overall RREH costs, for example, by using the oxygen byproduct from electrolysis to meet the demand of local hospitals. It may also consist in oversizing desalination capacity to benefit from economies of scale to provide water in water-scarce regions. Finally, this step may also highlight the need for new components, such as a water pipe to supply water to nearby farming installations.

Finally, Step 7 advises repeating the steps to refine the RREH design based on criteria such as local integration, number of required processes, or identifying better optimized designs. To model and optimize the RREH, the Graph Based Optimization Language (GBOML) introduced in \cite{miftari2022gboml} can be used. GBOML is specifically designed to solve optimization problems represented as graphs, making it well-suited for modeling the technological graph of an RREH.

\subsection{Systematic approach: an example}\label{subsec:example_prodcedure}

\begin{figure*}[h]
    \centering
    \includegraphics[width=0.9\textwidth]{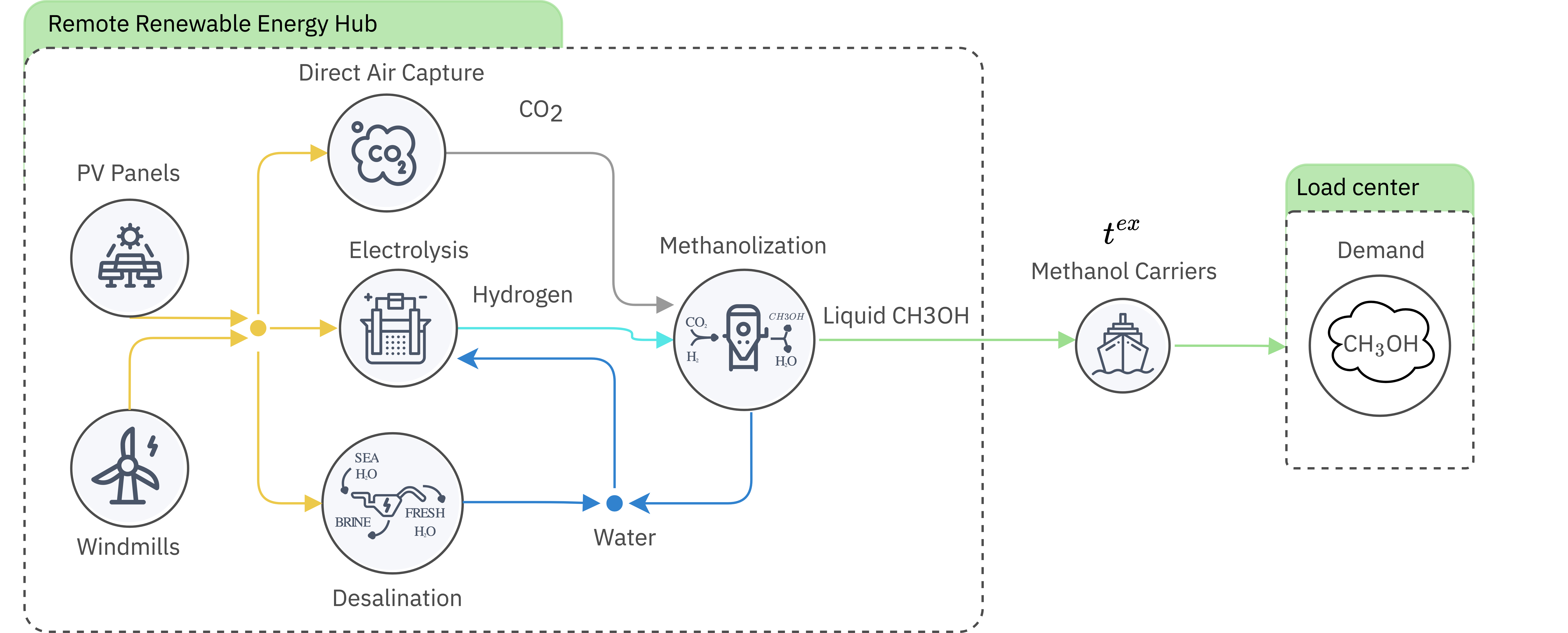}
    \caption{RREH located in Australia exporting CH$_3$OH to a load center situated in South Korea. }
    \label{fig:RREH_CH3OH}
\end{figure*}

The systematic approach proposed in \autoref{subsec:systematic_approach} is illustrated through an example inspired by \cite{BP_AustralianRenewableEnergyHub}. This example is for illustrative purposes only, as designing a new RREH requires significant effort and is beyond the scope of this section. Following the proposed approach, a hypothetical RREH design for methanol production and export to South Korea is considered. The design process follows these steps:

\begin{enumerate}

    \item \textbf{Define Export Commodities}: In this example, the RREH aims to export methanol. Thus, the export set is defined as: $\mathcal{E}_r = \{ CH_3OH \}$.
    
    \item \textbf{Select Locations}: Australia is a suitable candidate based on renewable energy potential from solar and wind resources and its viscinity to South Korea. Therefore, the set of locations is: 
    \[
    \mathcal{L}_r = \big\{ l = (\text{North Australia}, \text{high renewable potential}, \text{medium demand}) \big\}.
    \]
    
    \item \textbf{Construct Technological Graph}: The methanol synthesis process requires H$_2$ and CO$_2$. These can be sourced using an electrolyzer and a carbon capture technology, such as direct air capture (DAC). The electrolyzer requires H$_2$O and electricity, while DAC requires electricity and water. Since H$_2$O is scarce in Northern Australia, seawater desalination may be necessary which also requires electricity. The electricity for each process can be generated from a combination of solar panels and wind turbines. The full derivation of this technological graph $\mathcal{G}_r$ is given in \autoref{tab:taxonomy_CH3OH} while \autoref{fig:RREH_CH3OH} summarizes it. 
    
    \item \textbf{Consider Imports}: Some required molecules can be imported. For instance, CO$_2$ could be sourced from Southern Australia, where large emitters are located. Assume that this RREH relies solely on DAC, the set of imports writes as: $\mathcal{I}_r = \{ \text{sea water} \} $. 
    
    \item \textbf{Assess Byproducts}: The process generates heat and O$_2$. A portion of the heat can be reused in DAC, whereas O$_2$ has no direct reuse in this context. Thus, the set of byproducts is $\mathcal{B}_r = \{\text{heat}, O_2\}$. 
    
    \item \textbf{Identify Local Opportunities}: A local opportunity could involve partial utilization of the produced methanol, for instance, in machinery operating in nearby mines. Therefore, the set of locally exploited opportunities is defined as $\mathcal{O}_r = \{ CH_3OH \}$.
    
    \item \textbf{Optimize}: Define the constraints for each technology in the technological graph. Implement the system using a modeling language such as GBOML \cite{miftari2022gboml}, then optimize it. Finally, analyze the results to identify the main cost drivers and potential efficiency improvements.

\end{enumerate}

Based on the obtained results, the procedure may be iterated to refine the model. In subsequent iterations, several aspects could be explored, among them (i) incorporating batteries and storage technologies into the technological graph to manage power intermittency and facilitate the operation of must-run technologies, and (ii) introducing an import commodity to evaluate whether reducing reliance on DAC technology lowers costs.

\section{Conclusion} \label{sec:conclusion}

This article introduces a definition of the concept of RREHs and a taxonomy to characterize them. It also demonstrates the use of this taxonomy on two examples. This taxonomy can enhance communication within the scientific community and foster research on RREH integration and improved designs.

More specifically, its systematic approach to characterizing RREHs enables more effective comparisons and, if coupled with an optimization procedure, can help identify technologies and interconnections that minimize production costs. It can also help identify missing components of particular interest that contribute to building these RREHs, such as a heat network installation to recover part of the heat from the byproduct set.

While our taxonomy can already be exploited as it is, we believe there are still relevant avenues for enriching it. Although this taxonomy focuses on the qualitative comparison of technical components in RREHs, incorporating in the taxonomy financial aspects — such as different financing models or profit-sharing mechanisms — could broader its scope.

The taxonomy should be complemented by quantitative comparisons, which are necessary for a full comparison of different RREH proposals. Therefore, optimization and modeling techniques complement this taxonomy by enabling the derivation of quantitative values such as energy production costs or marginal costs of CO$_2$ captured, which are essential for comparing RREH projects.

Social and environmental indicators are also complementary to this taxonomy. These indicators can assess how an RREH project performs in achieving some of the United Nations (UN) Sustainable Development Goals, such as no poverty, decent work and economic growth, and climate action. Furthermore, one could incorporate these indicators into the objectives optimized during the systematic approach proposed to identify new designs.

Lastly, some external factors that are difficult to encompass within a formal framework such as the taxonomy are important to consider. For example, political stability in the region of the RREH can be a determining factor. Indeed, this external factor can significantly impact the financial costs of the project — higher risks may lead to increased borrowing costs — and influence the security of supply for load centers, which might be reluctant to sign long-term contracts.

\section*{Acknowledgements}

The authors thank Antoine Larbanois and Guillaume Derval for interesting discussions in the early stages of this research. Victor Dachet gratefully acknowledges the financial support of the Wallonia-Brussels Federation for his FRIA grant.

\section*{Competing interests}
The authors declare no competing interests.

\section*{Declaration of Generative AI and AI-assisted technologies in the writing process}
During the preparation of this work the author(s) used ChatGPT and Bard in order to correct the readiness, grammar and spelling of the writing. After using this tool/service, the authors reviewed and edited the content as needed and take full responsibility for the content of the publication.

\bibliography{bibli}

\appendix
\section{Example 2}

\begin{figure*}[h]
    \centering
    \includegraphics[width=0.9\textwidth]{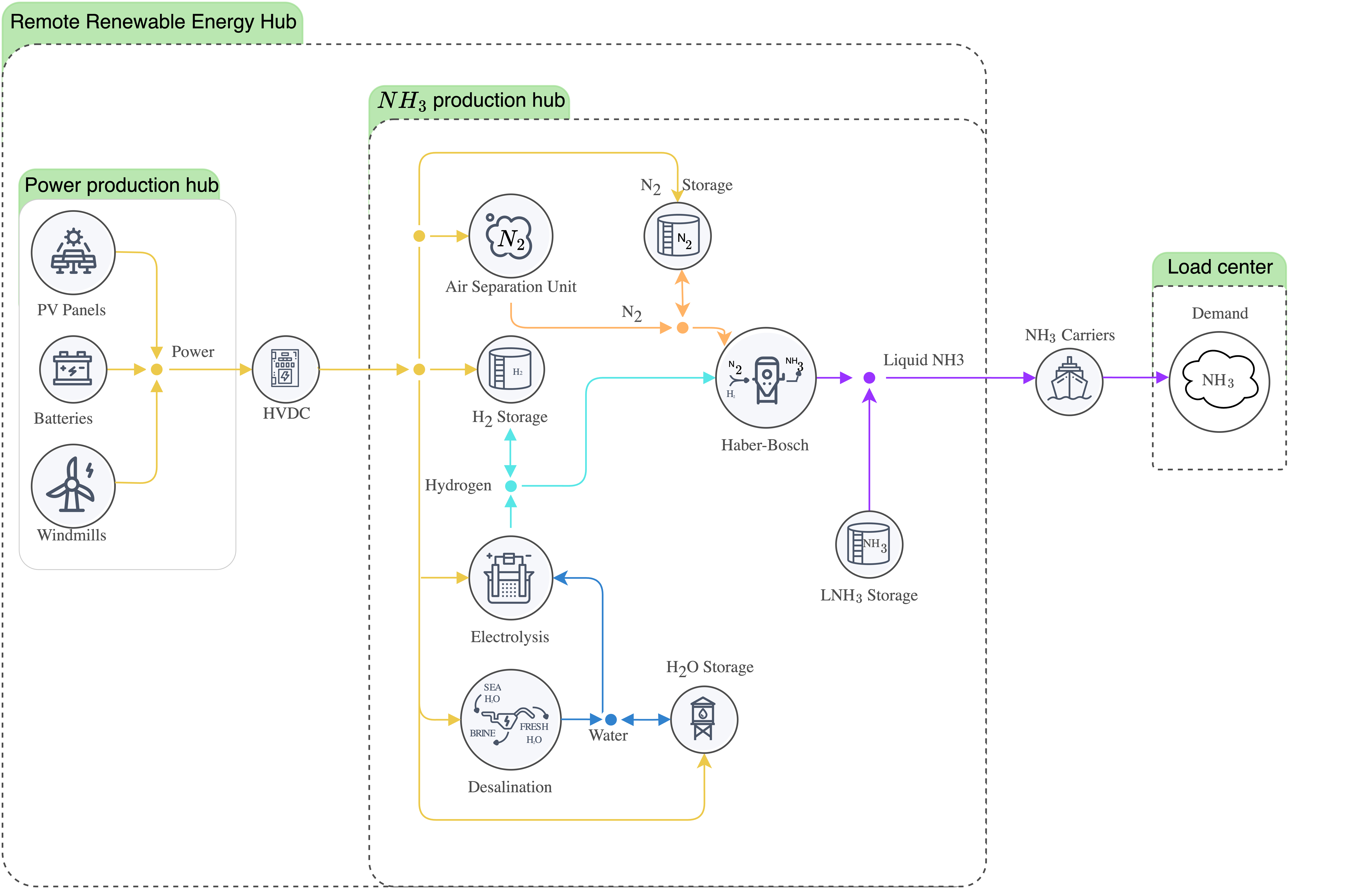}
    \caption{RREH located in Algeria exporting NH$_3$ to a load center situated in Belgium, adapted from \cite{Larbanois2023MultiCarriers}. }
    \label{fig:RREH_NH3}
\end{figure*}

\thispagestyle{empty}
\begin{table*}[htbp]
    \small
    \centering
    \caption{Expression of the full taxonomy for the second example in \autoref{sec:discussion}, where the technological graph and commodity flows $\mathcal{G}_{r}$ are defined by their sets of nodes and edges, $(\mathcal{T}_{r}, \mathcal{H}_{r})$. }
    \begin{tabular}{|p{2cm}|p{12cm}|}
        \hline
        \textbf{Set} & \textbf{Description} \\
        \hline
        $\mathcal{L}_r$ & \begin{description} 
            \item $\{ l_1, l_2, l_3\}$
            \item $\text{ where} $
            \item $l_1 = (\text{Sahara desert},  \text{high renewable potential}, \text{low demand}),$ 
            \item $l_2 = (\text{Algerian coast}, \text{high renewable potential}, \text{medium demand}) \},$ 
            \item $l_3 = (\text{From Sahara desert to the coast}, \text{high renewable potential}, \text{low demand})$
            
        \end{description} \\
        \hline
        $\mathcal{T}_{r}$ & \begin{description}
    \item $\{(PV_{l_1}, \{ \emptyset \}, \{\text{electricity} \}),$
    \item $(\text{Wind}_{l_1}, \{ \emptyset \}, \{\text{electricity} \}),$
    \item $(\text{Battery}_{l_1}, \{\text{electricity} \}, \{\text{electricity} \}),$
    \item $(\text{HVDC}_{l_3}, \{\text{electricity} \}, \{\text{electricity} \}),$
    \item $(\text{electrolyzer}_{l_2}, \{\text{electricity}, \text{H}_2\text{O} \}, \{\text{H}_2, \text{O}_2 \}),$
    \item $(\text{Desalination}_{l_2}, \{\text{electricity}, \text{sea water} \}, \{\text{H}_2\text{O} \}),$
    \item $(\text{H}_2\text{-Storage}_{l_2}, \{\text{electricity}, \text{H}_2 \}, \{\text{H}_2 \}),$
    \item $(\text{Air Separation Unit}_{l_2}, \{\text{electricity} \}, \{\text{N}_2, \text{Ar} \}),$
    \item $(\text{N}_2\text{-Storage}_{l_2}, \{\text{electricity}, \text{N}_2 \}, \{\text{N}_2 \}),$
    \item $(\text{Haber Bosch}_{l_2}, \{\text{N}_2, \text{H}_2 \}, \{\text{NH}_3 \}),$
    \item $(\text{NH}_3\text{-Storage}_{l_2}, \{\text{NH}_3 \}, \{\text{NH}_3 \}),$
    \item $(\text{export}_{l_2}, \{\text{NH}_3 \}, \{\emptyset \})\}$
\end{description} \\
        \hline
        $\mathcal{H}_{r}$ & 
            \begin{description}
                \item $\{(\textbf{Electricity}, \{\text{Wind}_{l_1}, \text{Battery}_{l_1}, PV_{l_1}\}, \{\text{Battery}_{l_1},\text{HVDC}_{l_1} \}),$ 
                \item $(\textbf{Electricity}, \{\text{HVDC}_{l_1}\}, $  
                \item \hspace{0.8cm} $\{\text{Battery}_{l_2},\text{electrolyzer}_{l_2}, \text{Desalination}_{l_2}, \text{H$_2$-Storage}_{l_2},DAC_{l_2},\text{CO$_2$-Storage}_{l_2}\}),$
                
                \item $(\textbf{H$_2$O}, \{\text{H$_2$-Storage}_{l_2}, \text{Desalination}_{l_2}, \text{Methanation}_{l_2} \},$
                
                \item \hspace{0.8cm} $\{ \text{H$_2$-Storage}_{l_2}, \text{electrolyzer}_{l_2} \}  ),$
                
                \item $(\textbf{H$_2$}, \{\text{H$_2$-Storage}_{l_2}, \text{electrolyzer}_{l_2} \}, \{\text{H$_2$-Storage}_{l_2}, \text{Haber Bosch}_{l_2} \}),$
                \item $(\textbf{NH$_3$}, \{ \text{Haber Bosch}_{l_2}, \text{NH$_3$-Storage}_{l_2} \}, \{ \text{NH$_3$-Storage}_{l_2}, t^{ex} \}),$
                \item $(\textbf{O$_2$}, \{ \text{electrolyzer}_{l_2} \}, \emptyset),$
                \item $(\textbf{Ar}, \{ \text{Air Separation Unit}_{l_2} \}, \emptyset),$
                \item $(\textbf{Heat}, \{ \text{electrolyzer}_{l_2} \}, \emptyset),$
                \item $(\textbf{Heat}, \{ \text{Haber-Bosch}_{l_2} \}, \emptyset),$
                \item 
            \end{description} \\
        \hline
        $\mathcal{C}_{r}$ & $ \{ \text{Electricity}, NH_3, H_2, H_2O, N_2, O_2, \text{Heat}, Ar \}$ \\
        \hline $\mathcal{E}_r $ & $ \{ NH_3 \}$  \\
        \hline $\mathcal{I}_r $ & $ \{ \text{sea water} \}$  \\
        \hline $\mathcal{B}_r $ & $ \{ O_2, \text{Heat}, Ar \} $ \\
        \hline $\mathcal{O}_r $ & $ \{  \} $ \\
        \hline
    \end{tabular}
    
    \label{tab:taxonomy-example2}
\end{table*}

\newpage
\section{Systematic approach example}

\begin{table}[h]
    \small
    \centering
    \caption{Expression of the full taxonomy for the example applying the systematic approach in \autoref{subsec:example_prodcedure}, where the technological graph and commodity flows $\mathcal{G}_{r}$ are described by their sets of nodes and edges $(\mathcal{T}_{r}, \mathcal{H}_{r})$. This RREH is designed to produce methanol in Australia for export to South Korea while utilizing part of the production domestically. }
    \begin{tabular}{|p{2cm}|p{10cm}|}
        \hline
        \textbf{Set} & \textbf{Description} \\
        \hline
        $\mathcal{L}_r$ & \begin{description} 
            \item $\{ l = (\text{North Australia}, \text{high renewable potential}, \text{medium demand}) \}$
            
        \end{description} \\
        \hline
        $\mathcal{T}_{r}$ & 
        
        \begin{description}
        \item $\{(PV_{l}, \{ \emptyset \}, \{\text{electricity} \}),$
        \item $(\text{Wind}_{l}, \{ \emptyset \}, \{\text{electricity} \}),$
        \item $(\text{electrolyzer}_{l}, \{\text{electricity}, \text{H}_2\text{O} \}, \{\text{H}_2, \text{O}_2 \}),$
        \item $(\text{Desalination}_{l}, \{\text{electricity}, \text{sea water} \}, \{\text{H}_2\text{O} \}),$
        \item $(\text{DAC}_{l}, \{\text{electricity}, \text{H}_2\text{O} \}, \{\text{CO}_2 \}),$
        \item $(\text{Methanolization}_{l}, \{\text{CO}_2, \text{H}_2 \}, \{\text{CH}_3\text{OH}, \text{H}_2\text{O} \}),$
        \item $(\text{export}_{l}, \{\text{CH}_3\text{OH}\}, \{\emptyset \})\}$
    \end{description} \\
        \hline
        $\mathcal{H}_{r}$ & 
            \begin{description}
                \item $\{(\text{Electricity}, \{\text{Wind}_{l}, PV_{l}\}, \{ \text{electrolyzer}_{l}, \text{Desalination}_{l} DAC_{l}\}),$ 
                \item $(\text{H$_2$O}, \{ \text{Desalination}_{l}, \text{Methanolization}_{l} \},\{\text{electrolyzer}_{l}, DAC_{l} \}  ),$
                \item $(\text{H$_2$}, \{\text{electrolyzer}_{l} \}, \{\text{Methanolization}_{l} \}),$
                \item $(\text{CH$_3$OH}, \{ \text{Methanolization}_{l} \}, \{ t^{ex} \}),$
                \item $(\text{O$_2$}, \{ \text{electrolyzer}_{l} \}, \emptyset),$
                \item $(\text{Heat}, \{ \text{electrolyzer}_{l} \}, \emptyset),$
                \item $(\text{Heat}, \{ \text{Methanolization}_{l} \}, \emptyset),$
            \end{description} \\
        \hline
        $\mathcal{C}_{r}$ & $ \{ \text{Electricity}, \text{CH}_3\text{OH}, \text{H}_2, H_2O, CO_2, O_2, \text{Heat} \}$ \\
        \hline $\mathcal{E}_r $ & $ \{ \text{CH}_3\text{OH} \}$  \\
        \hline $\mathcal{I}_r $ & $ \{ \text{sea water} \}$  \\
        \hline $\mathcal{B}_r $ & $ \{ O_2, \text{Heat} \} $ \\
        \hline $\mathcal{O}_r $ & $ \{ \text{CH}_3\text{OH} \} $ \\
        \hline
    \end{tabular}
    
    \label{tab:taxonomy_CH3OH}
\end{table}

\end{document}